\documentclass[12pt]{elsart}
\usepackage{epsfig,natbib}

\begin{document}
\runauthor{Ray and Bagla}
\begin{frontmatter}
\title{A Parallel TreePM Code}
\author{Suryadeep Ray},
\ead{surya@mri.ernet.in}
\author{J.S. Bagla}
\ead{jasjeet@mri.ernet.in}
\address{Harish-Chandra Research Institute, Chhatnag Road, Jhusi,
Allahabad-211019}

\begin{abstract}
We present an algorithm for parallelising the TreePM code.  
We use both functional and domain decompositions.  
Functional decomposition is used to separate the computation of long
range and short range forces, as well as the task of coordinating
communications between different components.
Short range force calculation is time consuming and benefits from
the use of domain decomposition.  
We have tested the code on a Linux cluster.
We get a speedup of $31.4$ for $128^3$ particle simulation on $33$ 
processors; speedup being  better for larger simulations. 
The time taken for one time step per particle is $6.5\mu$s for a
$256^3$ particle simulation on $65$ processors, thus a simulation that
runs for $4000$ time steps takes $5$ days on this cluster.
\end{abstract}


\end{frontmatter}


\section{Introduction}

Observations of large scale structures like galaxies, clusters of
galaxies along with observations of the cosmic microwave background
radiation (CMBR) can be put together in a consistent framework if we
assume that the large scale structures formed by gravitational
amplification of density perturbations
\citep{tp93,peebles,peacock,lss_review}.  
These perturbations had a very small amplitude at the time of
decoupling of matter and radiation, hence the highly isotropic
character of the CMBR.
Perturbations grow as overdense regions accrete mass and  galaxies form
when such regions are dense enough for star formation to take place.
Early evolution of perturbations can be studied analytically using
perturbation theory and approximation schemes.
A detailed study of non-linear evolution of density perturbations
requires the use of numerical simulations.
Several methods have been developed for simulating gravitational
clustering and formation of large scale structures, e.g. see
\citet{Bertsc98} for a review.
The main driving force for these developments has been the need to
simulate large systems in great detail while keeping errors in
control. 
The emergence of Beowulf clusters as an affordable platform for high
performance computing has given a fresh impetus to this activity, and
the focus has shifted to algorithms that can be parallelised easily on
such platforms
\citep{thesis,tpm,treepar96,tpmn,Vspr,mlapm,tpmnew,partreepm,gotpm,pmfast}.   
In this paper we present an algorithm for a parallel TreePM code.
The TreePM method~\citep{treepm,error_treepm} combines the tree
code~\citep{bh86} with a Particle-Mesh (PM) code, e.g. see
\citet{jbtp_pm,sim_book}. 
A brief summary of the TreePM method is given below,  we refer the reader to 
\citet{treepm}, and, \citet{error_treepm} for more details and comparison with
similar methods. 

Description of the TreePM method is followed by a discussion of the
parallelisms inherent in the algorithm.  
In later sections we proceed to discuss our implementation and the
performance.

\section{The TreePM method}

In the TreePM method the force computation is divided into two parts
by explicitly partitioning it into a long range and a short range
component.
Solution to the Poisson equation in Fourier space can be split
into two parts by partitioning of unity.  
This gives us the short range and the long range potential.
\begin{eqnarray}
\varphi_k &=& - \frac{4 \pi G \varrho_k}{k^2} \label{pm_std}\\
 &=& - \frac{4 \pi G \varrho_k}{k^2} \exp\left(-k^2 r_s^2\right)  -
 \frac{4 \pi G \varrho_k}{k^2} \left[1 - \exp\left(-k^2
 r_s^2\right)\right]\nonumber\\ 
 &=& \varphi_k^l + \varphi_k^s \nonumber \\ 
\varphi_k^l &=& - \frac{4 \pi G \varrho_k}{k^2} \exp\left(-k^2
 r_s^2\right) \label{longr}\\
\varphi_k^s &=& - \frac{4 \pi G \varrho_k}{k^2} \left[1 - \exp\left(-k^2
 r_s^2\right)\right] \label{shortr}
\end{eqnarray}
where $\varphi^l$ and $\varphi^s$ are the long range and the short range
potentials respectively.
$G$ is the gravitational coupling constant and $\varrho$ is density.
Here $r_s$ is the scale that is introduced to partition the
potential.  
From our earlier studies we found that the Gaussian is the best
partitioning and the optimum value for the scale $r_s$ is the
mean inter-particle separation~\citep{error_treepm}. 
The expression for the short range force in real space is:
\begin{equation}
{\bf f}^s({\bf r}) = - \frac{G m {\bf r}}{r^3} \left[{\rm
erfc}\left(\frac{r}{2 r_s}\right) + \frac{r}{r_s \sqrt{\pi}}
\exp\left(-\frac{r^2}{4 r_s^2}\right)\right] \label{fshort}
\end{equation}
Here, ${\rm erfc}$ is the complementary error function.
The long range potential is computed in the Fourier space, just as in
a PM code, but using eqn.(\ref{longr}) instead of eqn.(\ref{pm_std}).
This potential is then used to compute the long range force.  
The short range force is computed directly in real space using
eqn.(\ref{fshort}) instead of the inverse square force in the tree
method. 
The short range force falls rapidly at scales $r \gg r_s$, and hence
we need to take this into account only in a small region around each
particle.  
We define a scale $r_{cut}$ as the distance up to which we sum the
short range force, we use $r_{cut}=5r_s$~\citep{error_treepm}. 
With the choice of parameters mentioned here, we find that the error
in force is small over the entire range of scales.  
Unlike cosmological tree codes, the errors are relatively small even
for a homogeneous distribution of particles.
The CPU time per step varies very slowly with the level of clustering.

\section{Parallelisms in the Algorithm}

Hybrid nature of the TreePM method forces us to adopt a more involved
scheme for parallelisation as compared to the tree method. 
The tree method is used in the TreePM to calculate the short range
force, we start by reviewing a scheme for parallelising the tree code.

An inherent parallelism in all N-Body codes is that the force on
particles can be calculated concurrently. 
Barnes-Hut tree codes~\citep{bh86} divide the simulation volume into
cells and only a small subset of the details of particle distribution
in distant cells is needed for computing the force. 
Thus it is natural to divide the simulation volume into domains with
equal computational load and force on particles in a given domain can
be computed by one processor.
The simulation volume is bisected recursively along orthogonal
directions, each bisection is carried out in such a way that the
computational load is equal on both sides\citep{thesis,treepar96,Vspr}.
After $m$ bisections, the simulation volume is divided into $2^m$
domains -- all with equal computational load.  
These can now be assigned to different processors and calculation of
force can be carried out concurrently.
The process of domain decomposition adds some overhead, but it is
small compared to the gain due to parallelisation.
Of course, this overhead increases as we increase the number of processors for
domain decomposition. 
For long range forces like gravity, each processor needs information
from all the other processors and hence the number of communications
required is significant.
This can be a serious impediment for scaling
the code on distributed memory machines for a large number of
processors. 
This problem is less serious for the TreePM code as the short range
force calculation requires communications with a much smaller number
of processors.

The TreePM method splits force computation into two parts, the
long range and the short range force.
The method described above serves to compute the short range force.
The long range force can be computed concurrently on a processor not
involved in computation of short range force, this is another
parallelism inherent in the TreePM algorithm.  
We need to exploit these two parallelisms of the algorithm for a successful
implementation of the parallel TreePM code.
However the presence of two independent parallelisms makes the task
of load balancing somewhat nontrivial and gives rise to complexities
discussed below. 

Only a small fraction of CPU time is used for computing the long range force in
the sequential code.
Thus the number of processors used for the PM calculation can be much
smaller than the total number of processors being used, in fact only
one processor for the long range force calculation is sufficient for
most cases.  
An obvious problem that arises is that load balancing will be achieved
only for a specific number of processors and the load balancing will
be less than perfect for a smaller number of processors.
If the number of processors is larger the number required for optimum load 
balancing, then the processors doing the short range force will have to
wait.  
The situation can be remedied by spreading the task of long range
force calculation over more than one processor as the number of
processors ($N_{proc}$) increases. 

We now proceed to describe the detailed algorithm that we have adopted and
summarise various options that we considered at each step. 

\subsection{Short range force}

We use domain decomposition for computing the short range force as it is a
natural solution for dividing the task of force calculation.
Recursive orthogonal bisection is used to divide the simulation volume into
domains with an equal number of particles.
As long as the number of particles in each domain is sufficiently large, we
find that dividing the simulation volume into domains with equal number of
particles is sufficient for load balancing and we need not explicitly create
domains with equal computational load.
Bisection of the simulation volume is carried out in a method similar to that
outlined by \citet{thesis}. 
The Cartesian grid construct of message passing interface (MPI)
\citep{ref_mpi} is used for easy book keeping. 

A more tricky problem is communicating information about particles in
neighbouring domains for completing the calculation of short range force.  
The direct approach, which also ensures load balancing, is to request the
processors corresponding to neighbouring domains for the relevant
information \citep{thesis,treepar96}. 
A variant of this method is to send positions of particles near the
boundary of domains and seek the partial force information.
The problem with these approaches is that the communication and computation
overhead is significant. 
For a three dimensional simulation where each domain is larger than
$r_{cut}$, the number of communications is  restricted to nearest
neighbours amongst domains.
The number of nearest neighbours is never greater than $26$ but two point
communications with $26$ processors, per processor, can add a significant
amount of overhead. 
Some alternatives that we have tried are:
\begin{itemize}
\item
List non-local particles along with recursive bisection.  Thus the list of
non-local particles needed for calculating short range force is made at the
same time as domain decomposition.
\item
Send vertices of domain to the master node and request for a list of
non-local particles.
\item
Send positions of all particles to all processors, each processor isolates the
list of non-local particles that are needed.
\end{itemize}
The first option adds to the time before calculation of short range force can
commence, this nearly doubles the time take for domain decomposition though
there is no additional overhead beyond this. 
The second option, if it uses asynchronous communications, can be an
attractive solution in combination with a master node for coordinating
communications.  
The master node acts as a communication agent and it receives
positions of particles from all the domains and sends a list of
non-local particles needed for computation of short range force to
each domain. 
Thus the number of communications per node decreases to a few but this
is done at the cost of adding another processor.
We find that the last option listed here is the fastest of the three but adds
large overheads in terms of memory requirements for each processor.
As long as memory is not a limitation, this is the best option and we choose
this for our final implementation.

\subsection{Long range force}

Long range force is calculated using the PM method but using a different
kernel. 
We use FFTW~(http://www.fftw.org) for computing Fourier transforms in
this calculation.
The force is communicated to all the other nodes directly.

\subsection{Communications}

At the start of each time step, particle positions and velocities 
are {\sl gathered} by the origin node on the Cartesian grid. 
Every particle in any domain on the Cartesian grid carries
an identity tag so that one can trace the trajectory of each particle
in a simulation. 
{\sl MPI\_Reduce} is used within the Cartesian grid to communicate particle
identities to the origin node.
Particle positions are {\it{broadcast}} by this node to all the 
processors on the Cartesian grid as well as to the processor which
computes the long range force. 
Each node on the grid uses this information to shortlist particles that are
not local to the domain represented by the node but are needed to complete the
short range force calculation. 
The origin node initiates the process of domain decomposition.
Several communicators are constructed in order to exploit optimised global
communications for concurrent message passing between distinct subsets of
nodes. 
The node computing the long range force {\it{broadcasts}} the entire force
array at the end of the process. 
Each node retains the force for particles within the local domain by using
identity tags and discards the remaining array.  

\section{Performance of the parallel code}

Performance of parallel programs are measured in terms of speed up,
where speed up is defined as the time taken to run the program on a
single processor divided by the time taken to run the same program on
$N_{proc}$ processors.  
For a fully parallelisable problem, this should scale as $N_{proc}$.
However in problems where load balancing is not perfect, and inter-process
communication or computational overhead due to parallelisation is significant,
speed up is less than $N_{proc}$. 
Our aim here is to use optimise our algorithm to make speed up as close to
$N_{proc}$ as possible, especially for a reasonably large $N_{proc}$.
The speedup efficiency is the speedup divided by $N_{proc}$.

\begin{figure}
\epsfig{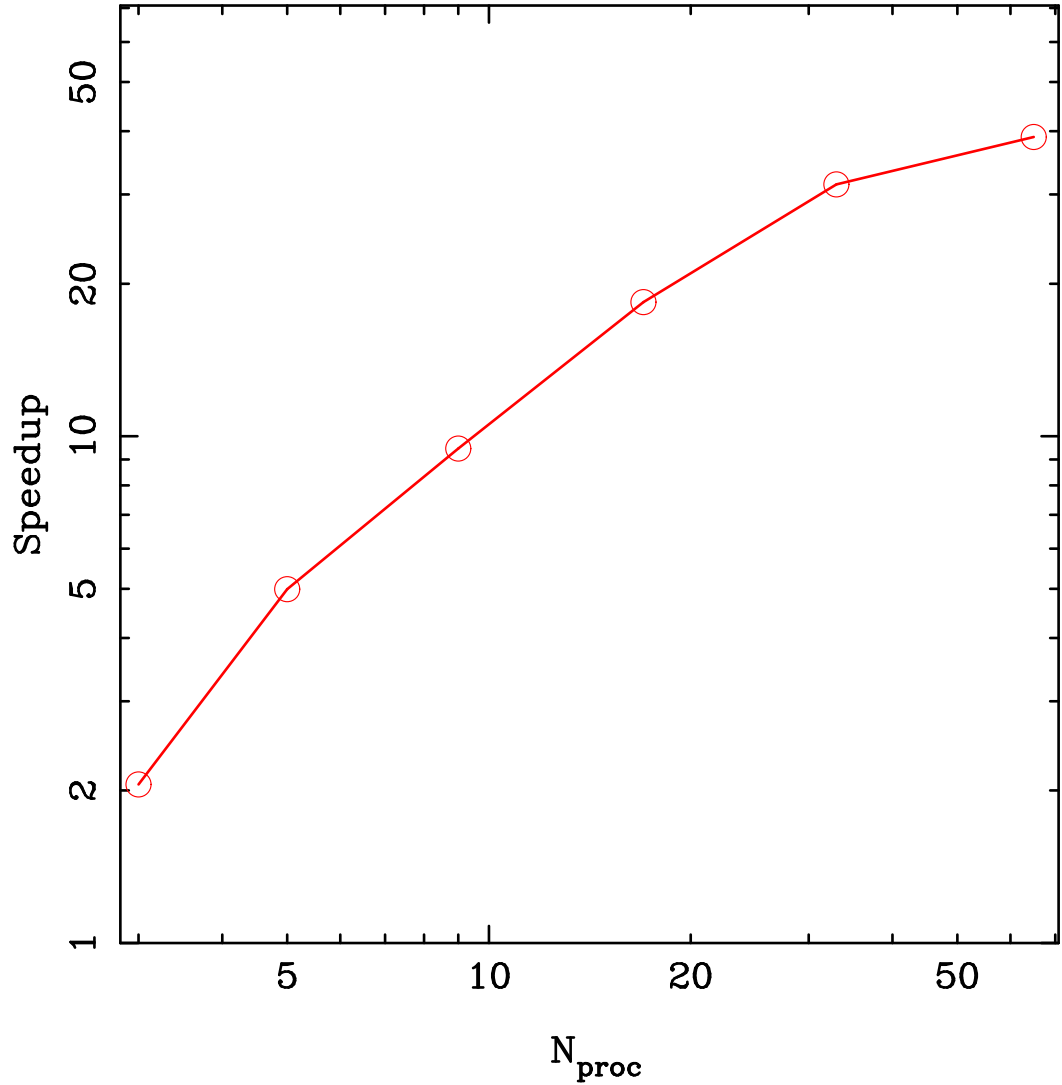}
\caption{The speedup is plotted here as a function of $N_{proc}$ for a $128^3$
  simulation (circles).  Features expected from the analysis of the algorithm
  are clearly seen here with the efficiency dropping off at both the small and
  the large $N_{proc}$, at large $N_{proc}$ the speedup begins to saturate and
  for small $N_{proc}$ the speedup decreases very rapidly.}  
\label{fig_speedup}
\end{figure}

\begin{figure}
\epsfig{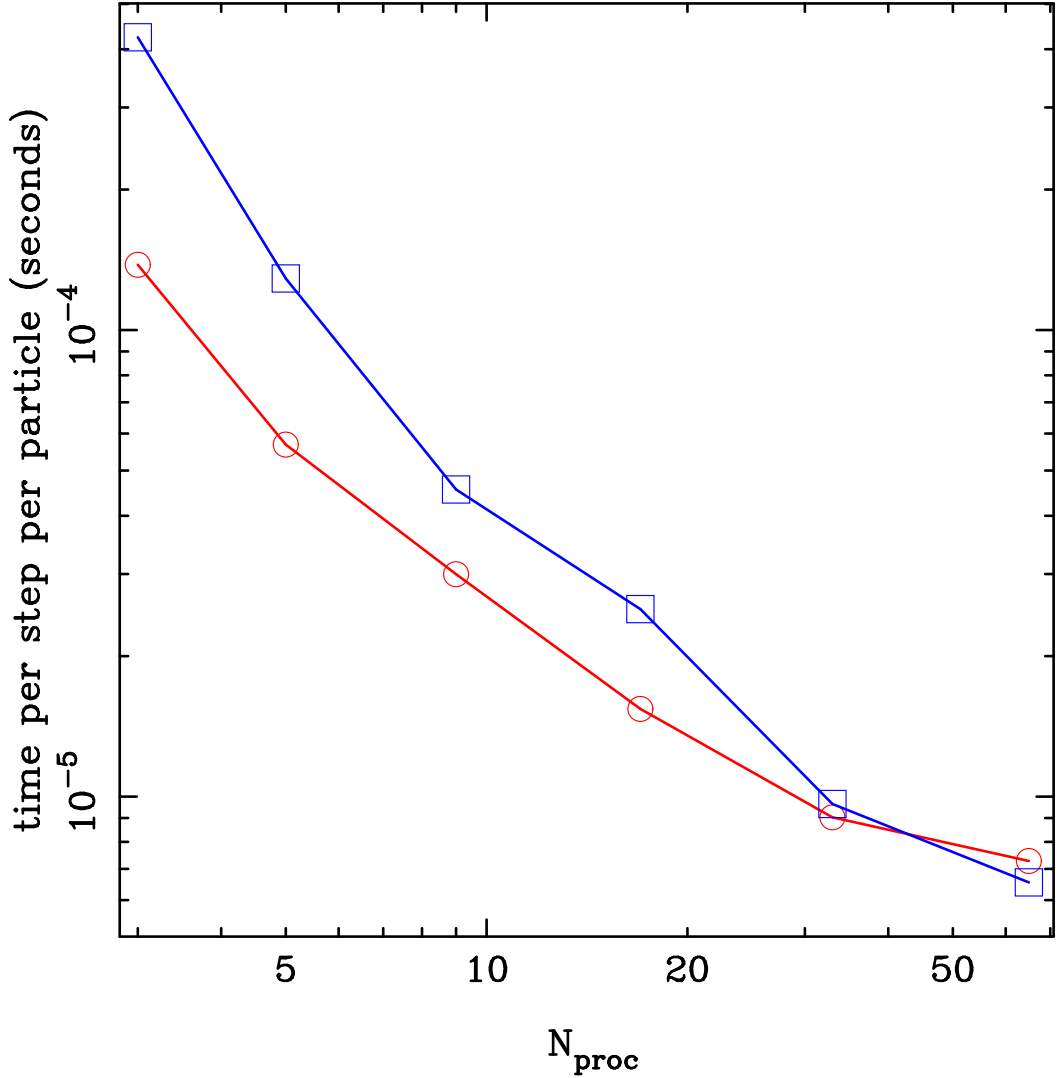}
\caption{The time taken per step per particle is plotted here as a function of
$N_{short}$ for simulations with $128^3$ (circles) and $256^3$ (squares)
particles.  Here we have used $N_{long}=1$, thus $N_{proc}=N_{short}+1$.  The
$256^3$ simulation requires about $6.5\mu$s per particle per time step for
$N_{proc}=65$.}
\label{fig_timing}
\end{figure}

If we use one processor for long range force calculation while changing
$N_{short}$, the number of processors computing the short range force, then
speedup will not be linear in $N_{proc}$.
For small $N_{short}$, the long range force calculation will take much
less time and the efficiency of parallelisation will be low due to
poor load balancing. 
As $N_{short}$ is increased, efficiency of parallelisation will improve till
load balancing is achieved.  
In the regime where $N_{short}$ is smaller than the optimum value for load
balancing, the code will speed up faster than linear.
For larger $N_{short}$, it will not be possible to load balance as
communication overhead and/or long range force calculation will take longer
than short range force calculation and there will no significant speed up.
The optimum value of $N_{short}$ depends on the size of the simulation and
details of how communications are organised.
These features can be seen in figure~\ref{fig_speedup} where the speedup
is plotted as a function of $N_{proc}$ for $128^3$ simulations. 
The speedup is almost linear beyond $N_{proc}=5$ for simulations with $128^3$
particles and it starts dropping beyond $N_{proc}=33$ and the speedup
efficiency falls below unity.
Data for this figure was obtained on a Linux cluster ({\sl Kabir},
see http://cluster.mri.ernet.in/ for details) with an SCI (scalable
coherent interface) network with computers connected along a $2$d torus.  
Each node is a dual processor workstation with $2.4$GHz Xeon processors. 
We obtain a speedup of $31.4$ on $33$ processors and $39$ on $65$ processors
for $128^3$ simulations.
Speedup is greater than the number of processors for $N_{proc}=9$ and $17$,
this can be explained in terms of improved cache performance for smaller data
sizes. 

Performance of the parallel code is presented in
fig.~\ref{fig_timing}, where we have plotted time taken per step per
particle as a function of $N_{proc}$. 
Notice that for $N_{proc}=65$, the time taken per step per particle is only
$6.5\mu$s for $256^3$ simulations, thus we can do a simulation of $4000$ time
steps in five days. 

We can make further improvement in our method by using more processors for
long range force calculation and by using a larger box for computing long
range force as this will reduce the communication overhead.
These changes, however, will be needed for a larger number of processors 
that we have access to.

\section{Discussion}

We have presented an algorithm for parallelising the TreePM code on a
Beowulf cluster.  
This code has been verified by comparing the final positions and velocities of
particles in some test cases with the output of the sequential code, therefore
the error profile of this code is same as the sequential TreePM code
\cite{error_treepm}. 
Even though we have tended to optimise the CPU time required at the cost of
memory requirements, the maximum memory requirement per node is about $80$
bytes times the number of particles for the double precision code.
We need up to $160$~MB per node for $128^3$ simulations and $1.25$~GB per node
for $256^3$ simulations. 
These numbers represent the maximum memory requirements and for much of the
time memory requirement is much smaller than this. 
Memory requirements can be reduced by about $25\%$ by reorganising the code
and adding a master node to gather positions and velocities of particles from
nodes that are calculating the short range force.

For $128^3$ simulations we get a speedup of $31.4$ on $33$ processors and $39$
on $65$ processors.  
The time taken for one time step per particle is $6.5\mu$s for a
$256^3$ particle simulation on $65$ processors, thus a simulation that
runs for $4000$ time steps takes $5$ days on this cluster.
These results are for a simulation with a global time step and further
optimisations in terms of individual time steps is being carried out.

The GOTPM code \cite{gotpm} has a better performance in terms of time taken
per particle per step.
Part of the speedup is due to use of a larger mesh for the long range force
calculation, and the remainder is due to a much smaller $r_{cut}$ and a more
relaxed cell acceptance criterion for calculation of the short range force.
The results for speedup efficiency and wall clock time per particle compare
well with the published numbers for other parallel N-Body simulation codes of
this class, e.g., \citet{Vspr,tpmnew}.

\section*{Acknowledgements}

The work reported here was done using the {\it Kabir} cluster at the
Harish-Chandra Research Institute (http://cluster.mri.ernet.in).

\end{document}